\documentclass[prmaterials,twocolumn,superscriptaddress,floatfix,showpacs,amsmath,amssymb]{revtex4}

\usepackage{epstopdf}
\usepackage{graphicx,color}
\usepackage{dcolumn}
\usepackage{bm}
\usepackage{amsmath}
\usepackage{mathtools}
\usepackage[latin1]{inputenc}
\usepackage{epstopdf}
\usepackage{subfigure}
\usepackage{textcomp}
\usepackage{booktabs,multirow}           
\newcommand{\ndb}{NdB$_4$}
\newcommand{\rb}{RB$_4$}

\newcommand{\tn}{$T_{\rm N}$}
\newcommand{\tione}{$T_{\rm IT}$}
\newcommand{\titwo}{$T_{\rm LT}$}

\newcommand{\gv}{$\gamma_{\rm v}$}
\newcommand{\gc}{$\gamma_{\rm c}$}
\newcommand{\gab}{$\gamma_{\rm \perp c}$}

\newcommand{\cp}{$c_{\rm p}$}
\newcommand{\cpm}{$c_{\rm p}^{\rm m}$}
\newcommand{\cpp}{$c_{\rm p}'$}

\newcommand{\jmk}{J/(mol\,K)}

\newcommand{\alv}{$\alpha_{\rm v}$}
\newcommand{\alm}{$\alpha_{\rm v}'$}

\begin{document}

\title{Magnetoelastic coupling and Gr\"{u}neisen scaling in NdB$_4$}

\author{R.~Ohlendorf}
\affiliation{Kirchhoff Institute for Physics, Heidelberg University, INF 227, D-69120 Heidelberg, Germany}
\email[Email:]{r.ohlendorf@kip.uni-heidelberg.de}
\author{S.~Spachmann}
\affiliation{Kirchhoff Institute for Physics, Heidelberg University, INF 227, D-69120 Heidelberg, Germany}
\author{L.~Fischer}
\affiliation{Kirchhoff Institute for Physics, Heidelberg University, INF 227, D-69120 Heidelberg, Germany}
\author{K.~Dey}
\affiliation{Kirchhoff Institute for Physics, Heidelberg University, INF 227, D-69120 Heidelberg, Germany}
\author{D.~Brunt}
\affiliation{Department of Physics, University of Warwick, Coventry, CV4 7AL, United Kingdom}
\author{G.~Balakrishnan}
\affiliation{Department of Physics, University of Warwick, Coventry, CV4 7AL, United Kingdom}
\author{O. A.~Petrenko}
\affiliation{Department of Physics, University of Warwick, Coventry, CV4 7AL, United Kingdom}
\author{R.~Klingeler}
\affiliation{Kirchhoff Institute for Physics, Heidelberg University, INF 227, D-69120 Heidelberg, Germany}
\affiliation{Centre for Advanced Materials (CAM), Heidelberg University, INF 225, D-69120 Heidelberg, Germany}
\email[Email:]{r.klingeler@kip.uni-heidelberg.de}



\date{\today}

\begin{abstract}
We report high-resolution capacitance dilatometry studies on the uniaxial length changes in a NdB$_4$ single crystal. The evolution of magnetically ordered phases below \tn\ = 17.2~K (commensurate antiferromagnetic phase, cAFM), \tione\ = 6.8~K (intermediate incommensurate phase, IT), and \titwo\ = 4.8~K (low-temperature phase, LT) is associated with pronounced anomalies in the thermal expansion coefficients. The data imply significant magneto-elastic coupling and evidence of a structural phase transition at \titwo . While both cAFM and LT favor structural anisotropy $\delta$ between in-plane and out-of-plane length changes, 
it competes with the IT-type of order, i.e., $\delta$ is suppressed in that phase. Notably, finite anisotropy well above \tn\ indicates short-range correlations which are, however, of neither cAFM, IT, nor LT-type. Gr\"{u}neisen analysis of the ratio of thermal expansion coefficient and specific heat enables the derivation of uniaxial as well as hydrostatic pressure dependencies. While $\alpha$/\cp\ evidences a single dominant energy scale in LT, our data imply precursory fluctuations of a competing phase in IT and cAFM, respectively. Our results suggest the presence of orbital degrees of freedom competing with cAFM and successive evolution of a magnetically and orbitally ordered ground state.
\end{abstract}

\maketitle

\section{Introduction}

Geometric frustration provides a prime route to macroscopic degeneracy of ground states, thereby suppressing or even completely eliminating static long-range order~\cite{Ram}. Often, when magnetic order eventually evolves, competing interactions result in complex magnetic phase diagrams constituting energetically similar spin configurations. Prototypical examples of geometrically frustrated magnets are triangular, kagome, and pyrochlore lattices~\cite{Shimizu,Shen,Pratt,Lee,Yan,Han,Gingras,Cha,Gar}. While many of these systems realize quantum magnets, rare earth (R) borides, due to large magnetic moment of the rare earth ions, are well suited for investigating frustrated magnets in a classical limit. The family of tetraborides RB$_4$ is one particular example exhibiting magnetic frustration due to the rare earth ions being arranged on the geometrically frustrated Shastry-Sutherland-lattice (SSL)~\cite{Sha,Gab}. Many of the \rb~compounds display a variety of competing phases at low magnetic fields~\cite{Gab}. As with all \rb~compounds featuring trivalent rare earth ions, \ndb\ reported here is metallic~\cite{Joh}. As such the RKKY interaction may play an important role in addition to competing interactions evolving on exchange paths formed by the SSL.

The \rb~family crystallizes in a tetragonal D$^5_{4h}$-P4/mbm structure~\cite{Eto}. The boron sublattice is made up of octahedra forming chains along the crystallographic $c$-direction. These chains are connected via two boron atoms to form rings within the $ab$-plane. The rare earth ions are located above and below the middle of these rings constituting the rare earth sublattice. In \ndb , magnetism is due to Nd$^{3+}$ ions which electron configuration 4$f^3$ implies a $\prescript{4}{}{I}_{9/2}$ ground state. Upon cooling, \ndb~shows three successive phase transitions signaling the onset of commensurate antiferromagnetic (cAFM) order at \tn\ = 17.2~K, an incommensurate antiferromagnetic phase (IT) evolving at \tione\ =~6.8~K and a low-temperature phase (LT) below \titwo\ =~4.8~K~\cite{Bru2, Wat}. The former phase features an all-in-structure of magnetic moments pointing into the squares formed by the Nd-ions~\cite{Met}. In addition, the moments are slightly tilted towards the $c$-axis~\cite{Met}. The pseudo-quartet ground state consisting of two Kramers doublets carries an electric quadrupole moment giving rise to orbital degrees of freedom~\cite{Yam}. Here, we study the effects of long-range order on the length changes in \ndb\ by high-resolution capacitance dilatometry. Despite the relevance of dilatometry for elucidating the coupling of lattice, orbital, and magnetic degrees of freedom particularly in rare-earth systems, such studies are yet missing for the whole family of tetraborides RB$_4$. We find pronounced thermal expansion anomalies at the magnetic transitions as well as evidence for a structural transition at \titwo . In addition, precursing fluctuations well above \tn\ associated with negative in-plane thermal expansion indicate short-range ordering in the paramagnetic phase which is, however, not associated with an ordering phenomena evolving in either of the three low-temperature phases. Analysis of the Gr\"{u}neisen  ratios and of structural anisotropy enables elucidating the interplay of lattice, structure, and presumingly orbital degrees of freedom.

\section{Experimental methods}

Single crystals of \ndb\ were grown by the optical floating-zone technique as reported in detail in Ref.~\onlinecite{Bru1}. The relative length changes $dL_i/L_i$ along the crystallographic [001] and [110] directions (space group 127), respectively, were studied on an oriented cuboid-shaped single crystal of dimensions $1.476 \times 1.478 \times 1.880~$mm$^{3}$. The measurements were done by means of a three-terminal high-resolution capacitance dilatometer~\cite{Kue2} and the linear thermal expansion coefficients ${\alpha_i=1/L_i\cdot dL_i(T)/dT}$ were derived. Static magnetic susceptibility $\chi=M/B$ was studied in a magnetic field of $B=0.1$~T in a vibrating sample magnetometer (VSM) of a Magnetic Properties Measurement System (MPMS3 SQUID magnetometer) by Quantum Design. Specific heat measurements have been performed in a Quantum Design PPMS using a relaxation method.

\section{Length changes upon evolution of magnetic order}

\begin{figure}[hb]
	\centering
	\includegraphics [width=0.95\columnwidth,clip] {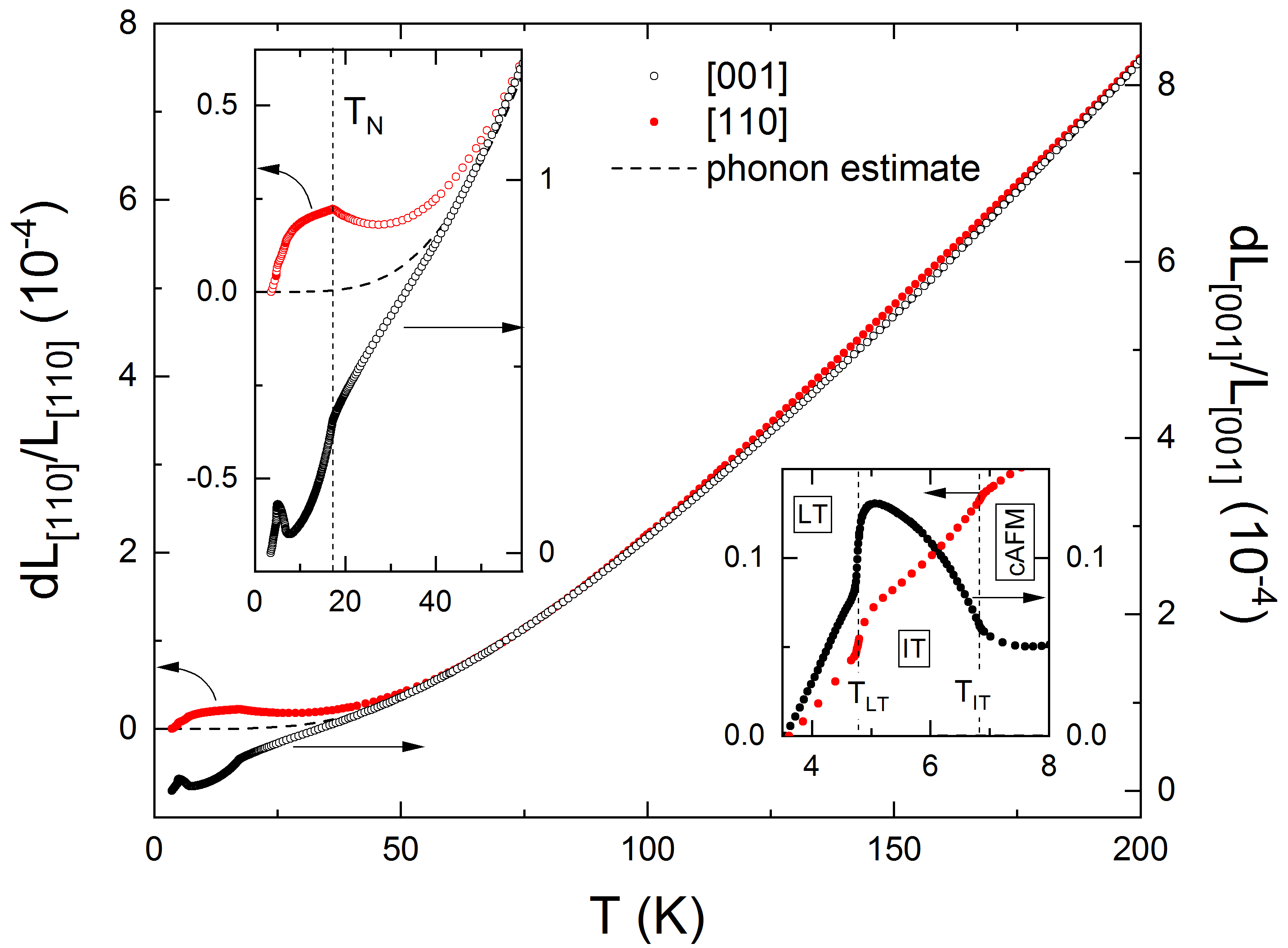}
	\caption{Relative length changes $dL_i/L_i$ along the crystallographic [001]- (black circles) and [110]-direction (red circles) $vs.$ temperature. Insets show enlargement of the low temperature region. Left and right ordinates are to scale but shifted in order to account for the difference at 200~K. Vertical dashed lines indicate the transition temperatures \tn , \tione , and \titwo\ towards the commensurate antiferromagnetic phase (cAFM), the incommensurate IT-phase, and the low-temperature phase LT.} \label{fig1}
\end{figure}

Thermal expansion in \ndb\ is positive at high temperatures and shows similar behavior in-plane and out-of-plane (see Fig.~\ref{fig1}). Upon cooling below $\simeq 50$~K, $L_{[001]}$ decreases significantly stronger as compared to $L_{[110]}$. Distinct anomalies in the thermal expansion at \tn\ as well as at \titwo\ and \tione\ imply pronounced magnetoelastic coupling.

\begin{figure}[htb]
	\centering
	\includegraphics [width=0.95\columnwidth,clip] {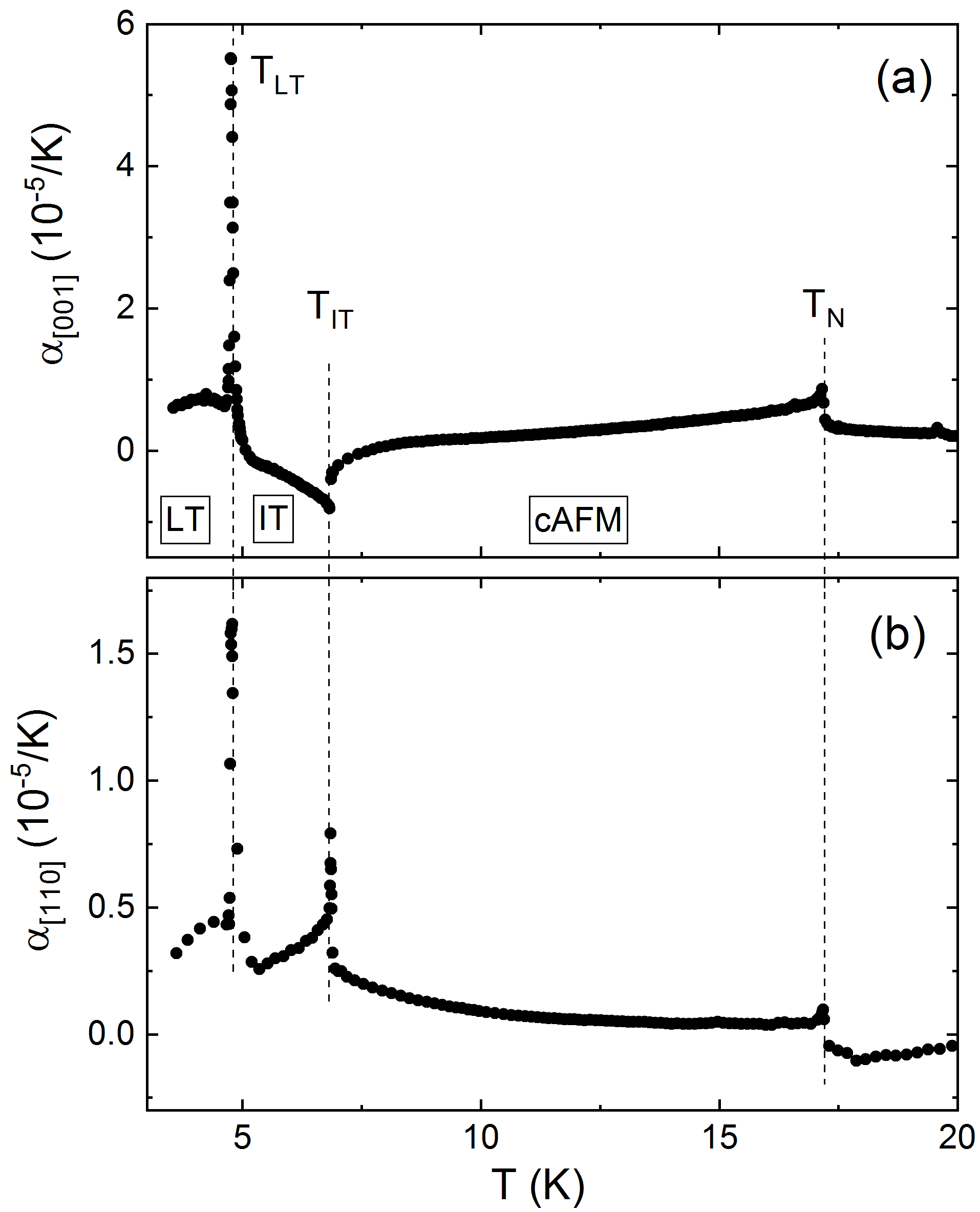}
	\caption{Uniaxial thermal expansion coefficients $\alpha_i$ along the crystallographic (a) [001]- and (b) [110]-direction. Dashed lines indicate the transition temperatures \titwo , \tione , and \tn\ (see the text).} \label{alphai}
\end{figure}

\begin{figure}[h]
	\centering
	\includegraphics [width=0.95\columnwidth,clip] {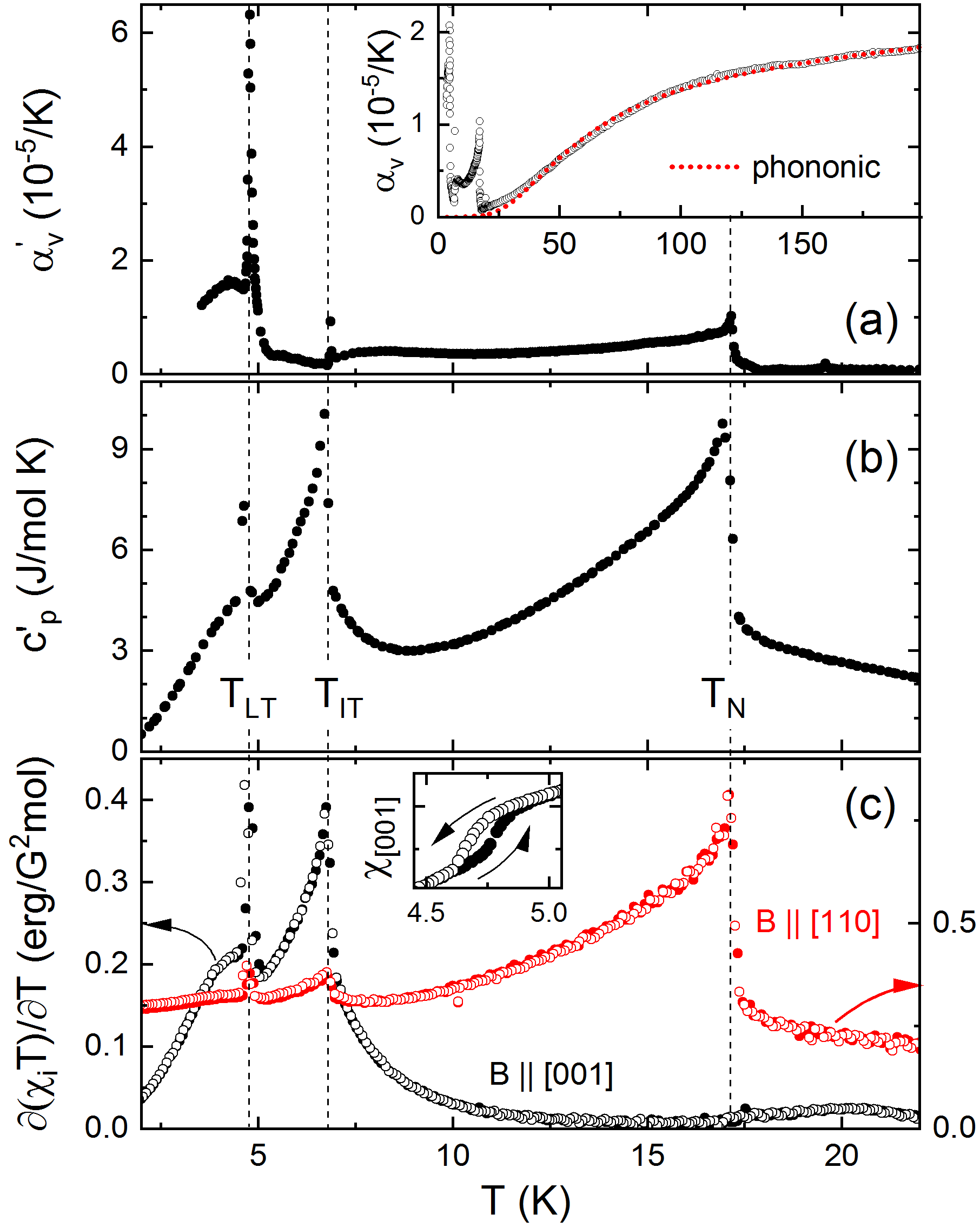}
	\caption{(a) Volumetric thermal expansion coefficient corrected by the phonon contribution. Inset: Uncorrected volumetric thermal expansion coefficient and phonon fit to the high temperature data (see the text). (b) Corresponding background corrected specific heat, and (c) Fisher's specific heat $\partial (\chi_i T)/ \partial T$ for $B=0.1$~T applied along the crystallographic [100]- (black circles) and [110]-direction (red circles), respectively. Filled (open) circles indicate measurements upon heating (cooling). Inset: Corresponding magnetic susceptibility $\chi_{[001]}$ highlighting hysteresis at \titwo . Vertical dashed lines indicate \titwo , \tione , and \tn .} \label{all}
\end{figure}

Magnetoelastic effects associated with the various ordering phenomena are particularly evident if the thermal expansion coefficients are considered, which display pronounced anomalies at \tn\ = 17.2~K, \tione\ = 6.8~K, and \titwo\ = 4.8~K (see Figs.~\ref{alphai} and \ref{all}a). The onset of long-range antiferromagnetic order at \tn\ is confirmed by previous magnetization and neutron diffraction data~\cite{Wat,Bru2,Met}, while two distinct antiferromagnetic phases develop at \titwo\ and \tione~\cite{Met}. At \tn\ and \tione , the thermal expansion data imply $\lambda$-shaped anomalies typical for  continuous phase transitions. Anomalies in the volume thermal expansion coefficient $\alpha_v=2\alpha_{\perp c}+\alpha_{\parallel c}$ show similar features (Fig.~\ref{all}a). All anomalies at \tn\ imply the decrease in length upon onset of magnetic order, i.e., there is positive uniaxial pressure dependence $dT_N/dp_i>0$ for both $i =$~[001] and [110]. Accordingly, hydrostatic pressure dependence of \tn\ is positive, too.

In contrast, anomalies in the uniaxial thermal expansion coefficients exhibit opposite signs at \tione , i.e., $dT_{\rm IT}/dp_{\parallel c}<0$ and $dT_{\rm IT}/dp_{\perp c}>0$ (Fig.~\ref{alphai}). The resulting volume response (Fig.~\ref{all}a) exhibits a positive sign of anomaly and hence positive hydrostatic pressure dependence. At \titwo\ there are jumps $\Delta L_i$ in the observed length changes for both directions $i$, which imply discontinuous evolution of the LT phase. The corresponding jumps in $L_i(T)$ are negative upon cooling (see the insets of Fig.~\ref{fig1}) which  implies $dT_{\rm LT}/dp_i>0$ in both cases. Quantitatively, at \titwo , we observe ${\Delta L_{\rm [110]}=-2.0(2)\cdot 10^{-6}}$ and ${\Delta L_{\rm [001]}=-5.0(2)\cdot 10^{-6}}$.

In order to quantitatively evaluate the effect of magnetic order and crystal field on thermal expansion, phonon contributions are estimated from specific heat data on LuB$_4$~[\onlinecite{Nov}] by assuming Gr\"{u}neisen scaling of the phonon heat capacity. Specifically, the phonon specific heat of LuB$_4$ is given in~\cite{Nov} as

\begin{align}
c_p^{ph}=& a_1D\left(\frac{T}{\Theta_{D1}}\right)+a_2D\left(\frac{T}{\Theta_{D2}}\right)+a_3E\left(\frac{T}{\Theta_{E1}}\right)\nonumber\\
&+a_4E\left(\frac{T}{\Theta_{E2}}\right).
\label{eq2}
\end{align}

Here, $a_i$ are parameters, and $D$ and $E$ denote Debye and Einstein functions with Debye and Einstein temperatures as given in Table~\ref{tabgruen}. The same function was fitted to the uniaxial thermal expansion coefficients $\alpha_{\parallel c}$ and $\alpha_{\perp c}$, above 65~K, using the same Debye and Einstein temperatures and allowing the parameters $a_i$ to vary. We note that, in LuB$_4$, a linear term for the electronic part is very small~[\onlinecite{Nov}] and has been omitted here. The volumetric thermal expansion coefficient obtained by $\alpha_v=2\alpha_{\perp c}+\alpha_{\parallel c}$ was fitted in the same way, except that $a_2$ was set to the value obtained from the fit to $\alpha_{\parallel c}$. From these fits the phononic Gr\"{u}neisen parameters $\gamma=\alpha /c_p$ are determined for both uniaxial thermal expansion coefficients and for the volume as listed in Table~\ref{tabgruen}.

\renewcommand{\arraystretch}{1.5}

\begin{table}[h!]
\setlength{\tabcolsep}{6pt}
\caption{Parameters of phonon background of thermal expansion. $\Theta_{\rm D}$ and $\Theta_{\rm E}$ are Debye- and Einstein-temperatures describing a phonon background of specific heat in LuB$_4$ from Ref.~\onlinecite{Nov}. The parameters $\gamma$ are the uniaxial and volume phononic Gr\"{u}neisen parameters.}
\label{tabgruen}
\begin{center}
\begin{tabular}{l | l l l}
		
\hline\hline
& &  $\gamma$ (mol/MJ) & \\
\hline
$\Theta_{\rm D1}=1140$~K & $\gamma_{\rm D1}^{\parallel c}$=0.028  & $\gamma_{\rm D1}^{\perp c}$=0.019  & $\gamma_{\rm D1}^{v}$=0.069 \\
$\Theta_{\rm D2}=211$~K  & $\gamma_{\rm D2}^{\parallel c}$=0.317  & $\gamma_{\rm D2}^{\perp c}$=0      & $\gamma_{\rm D2}^{v}$=0.317 \\
$\Theta_{\rm E1}=130$~K  & $\gamma_{\rm E1}^{\parallel c}$=0      & $\gamma_{\rm E1}^{\perp c}$=0      & $\gamma_{\rm E1}^{v}$=0 \\
$\Theta_{\rm E2}=190$~K  & $\gamma_{\rm E2}^{\parallel c}$=0.356  & $\gamma_{\rm E2}^{\perp c}$=0.680  & $\gamma_{\rm E2}^{v}$=1.694 \\

\hline\hline
	
\end{tabular}	
\end{center}
\end{table}

\renewcommand{\arraystretch}{1}


The result of this estimate shown by the dashed line in Fig.~\ref{fig1} implies anisotropic non-phononic length changes up to about 65~K while for higher temperatures the experimental data are well described by the fits (see Fig.~\ref{fig1}).


Figure~\ref{all}a shows the volumetric thermal expansion coefficient \alm\ corrected by the phonon contribution. The inset shows the bare experimental data \alv\ and the phonon fit derived as described above. In the whole temperature regime under study, \alm\ is positive and in particular -- as described above -- all anomalies are positive indicating positive hydrostatic pressure dependencies $dT_i/dp>0$ of all transition temperatures. In addition, Fig.~\ref{all}b displays the background corrected specific heat \cpp\ obtained by analogously subtracting the phonon entropy changes. For comparison, we also present Fisher's specific heat~\cite{Fis} $c_{\rm p}^{\rm m}\propto \partial(\chi_{i} T)/\partial T$ for magnetic fields applied $B||c$ and $B\perp c$ (Fig.~\ref{all}c) which provides an estimate of magnetic entropy changes. Analogous to $\alpha$, \cpp\ and \cpm\ display $\lambda$-shaped and discontinuous anomalies at the respective ordering temperatures. Note that in general, \cpm\ resembles \cpp\ very well, which strongly confirms the procedure applied for background correction. The fact that there is no visible anomaly in $\partial (\chi_{c}T)/\partial T$ at \tn\ implies the easy axis being $\perp c$ in the commensurate antiferromagnetic phase (cAFM) which agrees to recent neutron data~\cite{Yam,Met}. In addition, the magnetic entropy changes are well detected by $\chi_{\| c}$  in the IT and LT phases, suggesting significant out-of-plane spin components. We also note a small but finite hysteresis between cooling and heating in the magnetic susceptibility associated with the discontinuity at \titwo\ (see the inset of Fig.~\ref{all}c).

\section{Discussion}

\begin{figure}[htb]
	\centering
	\includegraphics [width=0.95\columnwidth,clip] {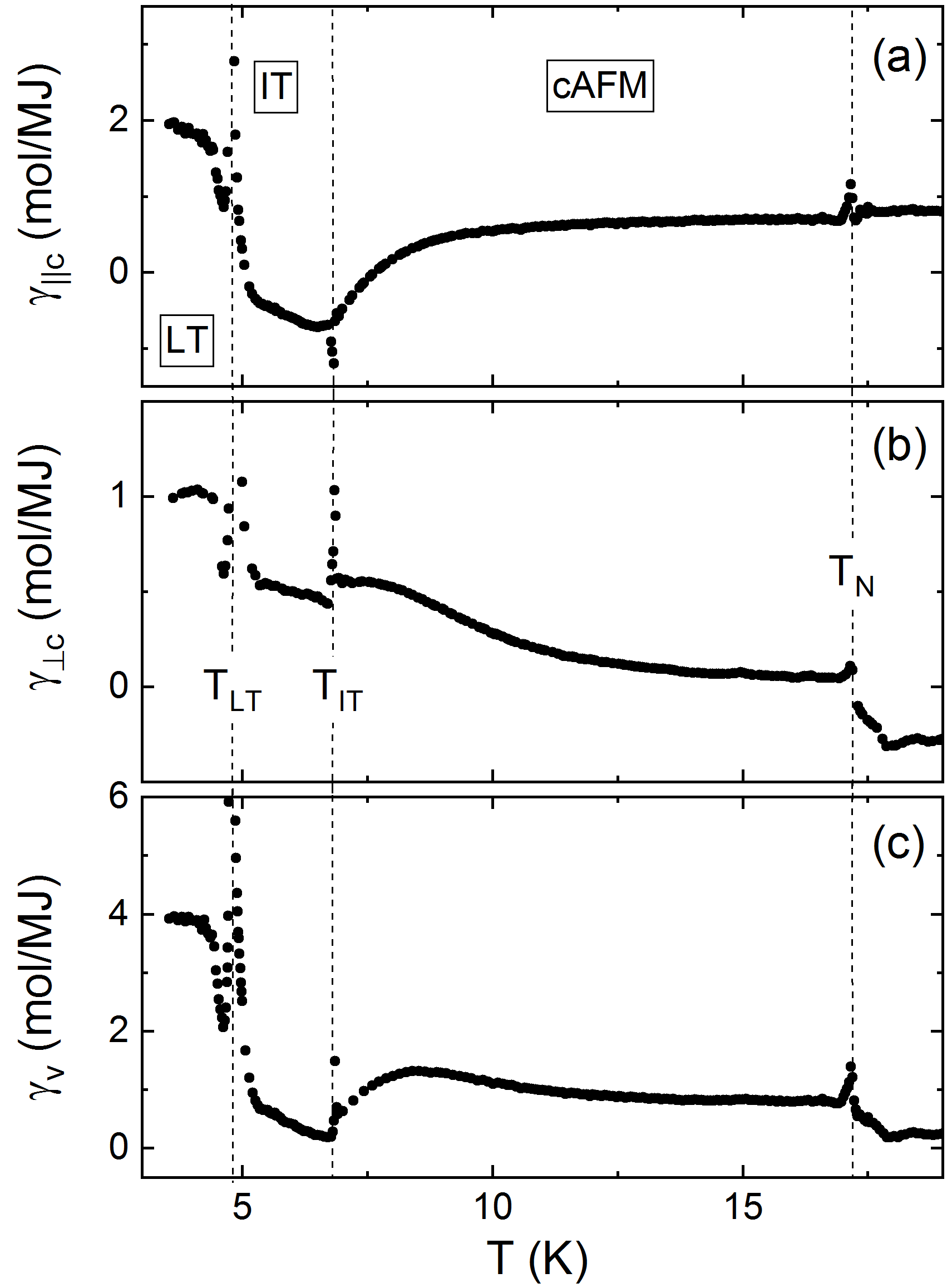}
	\caption{Gr\"{u}neisen parameters $\gamma_i=\alpha_{i}'$/\cpp\ where $i$ denotes the two directions [001] (a) and [110] (b) as well as the volume (c). Dashed lines show the transition temperatures \tn , \tione , and \titwo .} \label{gruen}
\end{figure}

Comparing the anomalous length and entropy changes provides quantitative information on coupling between spin and structure. In the presence of one dominant energy scale $\epsilon$, the ratio of the thermal expansion coefficient and specific heat is independent on temperature and enables the determination of the pressure dependence of $\epsilon$ by evaluating the volume or uniaxial Gr\"{u}neisen parameter~\cite{Gegenwart,Klingeler}

\begin{equation}
\gamma_{i} = \frac{\alpha_{i}}{c_{\rm p}} = \frac{1}{V_{\rm m}}\left. \frac{\partial \ln \epsilon}{dp_{i}}\right|_T . \label{eqgruen}
\end{equation}

Here, $V_{\rm m}$ is the molar volume and the index $i$ indicates a uniaxial direction or the volume. Exploiting the non-phononic contributions $\alpha_{\rm [110]}'$, $\alpha_{\rm [001]}'$, \alm , and \cpp\ yields the Gr\"{u}neisen parameters in Fig.~\ref{gruen}. The experimentally obtained Gr\"{u}neisen ratios distinctly change at the different phase boundaries thereby signaling clear changes of magnetoelastic coupling. At \tn\ we observe jumps in \gv\ and \gab\ but no clear anomaly appears in \gc . This absence of a clear anomaly corresponds to the absence of an anomaly in the magnetic susceptibility $\chi_{\rm [001]}$ at \tn . Within the cAFM phase, both the uniaxial Gr\"{u}neisen parameters $\gamma_{\parallel c}$ and $\gamma_{\perp c}$ as well as the volume Gr\"{u}neisen parameter $\gamma_v$ are rather constant but begin to change a few Kelvin above \tione . We conclude that critical fluctuations start to evolve at around 10~K, associated with structural changes of opposite uniaxial pressure dependence. The presence of opposite signs of anomalies tends to volume conservation at the phase transition and is also a feature of the thermal expansion anomalies at \tione\ (see Fig.~\ref{alphai}). We hence conclude that fluctuations below 10~K are of the IT-type. Notably, the changes of Gr\"{u}neisen scaling in the cAFM phase correspond to the evolution of the purely magnetic (1 0 0) reflection which gradually evolves below around 10~K and sharply increases at \titwo\ \cite{Met}. Microscopically, the  fluctuation regime in cAFM seems to be associated with the rotation of magnetic moments out of the $ab$-plane as the out-of-plane angle is found to evolve below ca.~10 K as a secondary order parameter. In the IT-phase, all parameters $\gamma_{i}$ increase upon cooling which we again attribute to precursing fluctuations -- here of the LT-type as corroborated by the equal signs of associated pressure dependencies -- superimposed to the response of the IT-type of order. Lastly there is a (non-volume conserving) jump at the transition into the LT-phase which features constant values of $\gamma_{i}$.

Our analysis yields the $\gamma$ values summarized in Table~\ref{tb1}. Using the Ehrenfest relation for the continuous phase transitions at \tn\ and \tione , the obtained values of $\gamma$ yield the pressure dependencies of the ordering temperatures at vanishing pressure, i.e., $dT_{\rm N/IT}/dp_i = T_{\rm N/IT} V_{\rm m}\gamma_i$.~\cite{Klingeler} Here, the pressure dependence of \tione\ is estimated by the value of the Gr\"{u}neisen parameter directly below the phase transition. For the discontinuous phase transition at \titwo , the pressure dependence is governed by the respective discontinuities in length, $\Delta L_{i,{\rm LT}}/L_i$ and in entropy, $\Delta S_{\rm LT}$, via the Clausius-Clapeyron equation:

\begin{equation}
\frac{dT_{\rm LT}}{dp_i}=V_m\frac{\Delta L_{i,{\rm LT}}/L_i}{\Delta S_{\rm LT}},
\end{equation}

with the molar volume $V_m = 3.22\cdot 10^{-5}$~m$^3$~\cite{Bru1}. Analysis of the experimental specific heat data yields $\Delta S_{\rm LT}=0.13(3)$~\jmk . We note a general experimental uncertainty due to the relaxation method used for determining the heat capacity at a discontinuous phase transition. From the field dependence ${dT_{\rm LT}/dB \approx 0.4(1)}$~K/T obtained around 1~T and the jump in magnetization $\Delta M_{\rm LT}\approx 85(10)$~erg/(mol$\cdot$G) we obtain $\Delta S_{\rm LT}= (dT_{\rm LT}/dB)^{-1}\cdot\Delta M_{\rm LT}\approx 0.21(8)$~\jmk\ which is hence used for the Clausius-Clapeyron analysis. The results for all phase transitions are shown in Table~\ref{tb1}.

\renewcommand{\arraystretch}{1.5}
\begin{table*}[htb]
\setlength{\tabcolsep}{15pt}
\caption{Pressure dependence of ordering temperatures and Gr\"{u}neisen parameter for the respective phases for uniaxial pressure along the [001]- and the [110]-direction as well as for hydrostatic pressure. The asterisks mark averaged values for the regimes where $\gamma$ varies. Parameters for the cAFM phase are obtained above 10~K.}
\label{tb1}
\begin{center}
\begin{tabular}{c | c c c c c c}
		
\hline\hline

& $dT/dp_{\rm c}$ & $\gamma_{\rm c}$ &$dT/dp_{\rm \perp c}$ & $\gamma_{\rm \perp c}$  &$dT/dp$    & $\gamma_{\rm v}$\\
&(K/GPa)                & (mol/MJ)               &(K/GPa)           &(mol/MJ)             &(K/GPa)    &(mol/MJ)\\
\hline
\tn\   &  0.38(4)          & 0.69(7)                & 0.034(8)         & 0.061(14)           & 0.44(19)  & 0.80(3)\\
\tione\ & -0.15*           & -0.69*                  & 0.11*        & 0.52*             & 0.04*      & 0.18* \\			
\titwo\ &  0.8(3)        &1.8(2)            & 0.3(1)         &1.0(1)              &1.3(6)    & 3.8(1)\\
\hline\hline
		
\end{tabular}	
\end{center}
\end{table*}
\renewcommand{\arraystretch}{1}

\begin{figure}[tb]
	\centering
	\includegraphics [width=0.95\columnwidth,clip] {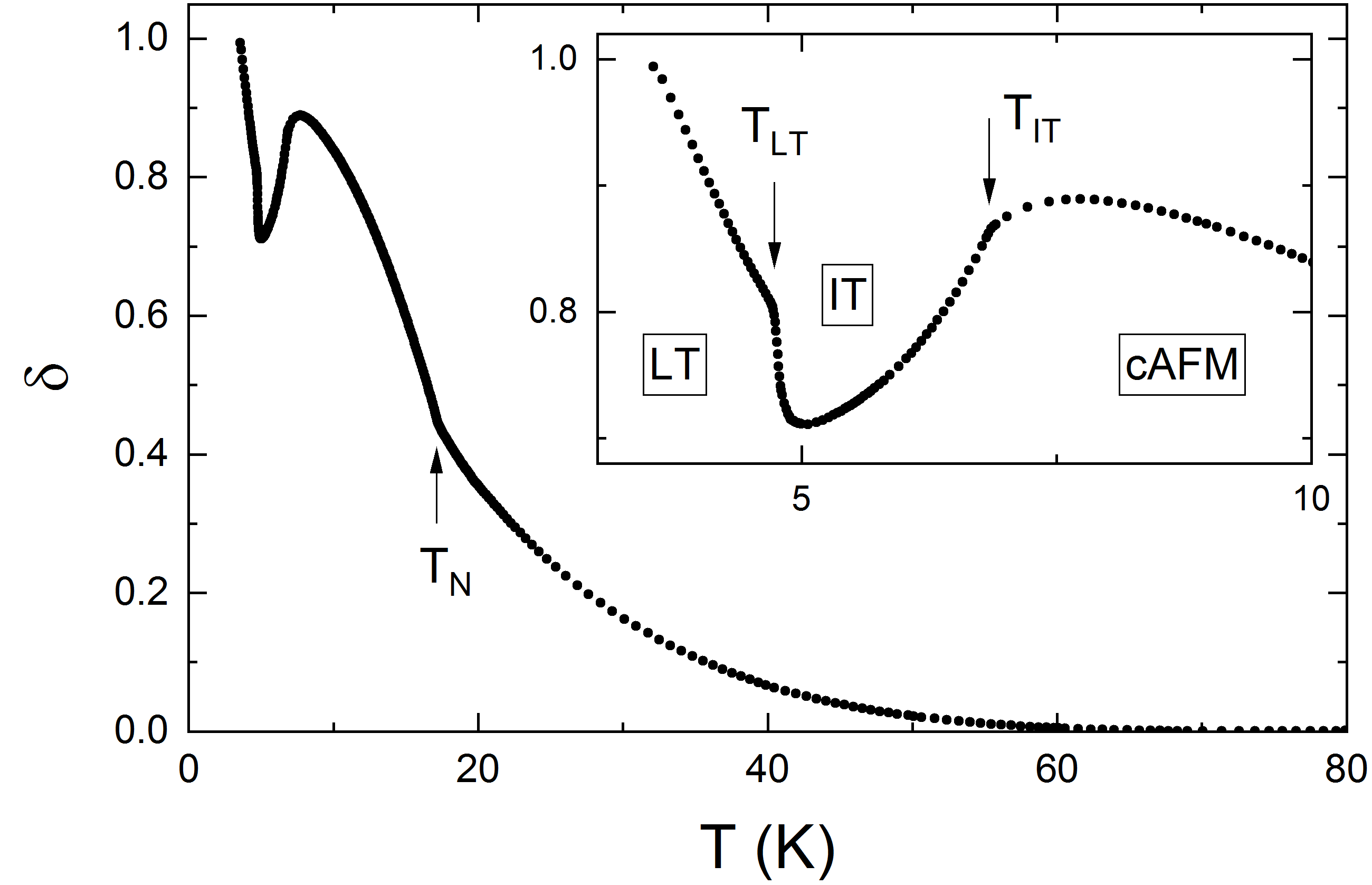}
	\caption{Temperature dependence of the relative distortion of the uniaxial length changes $\delta = (dL_{[110]}-dL_{[001]})$/$(dL_{[110]}+dL_{[001]})$. The inset shows an enlargement of the low-temperature behavior. Arrows indicate the transition temperatures of competing magnetic phases.} \label{OP}
\end{figure}

The thermal expansion data do not show clear signatures associated with the CF levels which have been detected at 2.9, 12.5 and 17.7 meV~\cite{Bru3}. Excitation of these CF levels would imply Schottky contributions to the thermal expansion in the paramagnetic phase centered around 60~K and 82~K, respectively, while a potential Schottky peak at around 14~K associated with the lowest CF level might be masked and shifted due to the presence of long-range magnetic order. The absence of clear anomalies above \tn\ is, however, evident from Fig.~\ref{fig1} and the inset of Fig.~\ref{all}a. Below \tn , there is no clear signature of CF levels either, which might have been also expected in the experimental Gr\"uneisen parameters in Fig.~\ref{gruen}~\cite{Werner2019}. As the magnitude and the shape of the Schottky anomalies in the thermal expansion coefficients are completely determined by the energy gaps $\Delta_i$ and their pressure dependencies $\partial \ln{\Delta}/\partial p_i$ (see Eq.~\ref{eqgruen}), we conclude rather small pressure dependencies of all $\Delta_i$ for hydrostatic pressure and uniaxial pressure along [110] and [001] (cf., e.g., \cite{Berggold2007}).

In order to highlight the difference of in-plane and out-of-plane behavior, a structural anisotropy parameter $\delta = (dL_{[110]}-dL_{[001]})$/$(dL_{[110]}+dL_{[001]})$ is presented in Fig.~\ref{OP}. It represents the difference of in-plane and out-of-plane length changes, assuming finite values below $T^{\ast}\sim 65$~K and smoothly increasing cubically upon cooling to \tn . Kinks at \tn\ and \tione\ as well as a jump at \titwo\ indicate the magnetic phase transition. While a rather conventional increase towards saturation is observed in cAFM, the distortion actually decreases in IT but eventually linearly increases in LT down to lowest accessible temperatures. The jump in length at \titwo\ implies that there is a combined structural-magnetic phase transition associated with abrupt shrinking of both $L_{\rm [001]}$ and $L_{\rm [110]}$, a jump-like increase of the structural anisotropy parameter $\delta$, and evolution of the LT-type of spin configuration~\cite{Bru3,Met}. 

The data in Fig.~\ref{OP} imply that commensurate magnetic order in HT and LT promotes structural anisotropy while in contrast $\delta$ competes with incommensurate magnetic order in IT. This competition is also evidenced by the Gr\"{u}neisen parameters which, in particular, display a sign change of $\gamma_{\perp c}$ at \tione\ and \titwo . We conclude the presence and interplay of several degrees of freedom. It is straightforward to attribute this observation to the magnetic and orbital degrees of freedom of the system. This leads to the scenario that the LT phase features combined magnetic and orbital order which evolution is associated with the qualitative change from a volume-conserving behaviour in the IT phase to a non-volume-conserving one below \titwo . Note, that only the magnetically incommensurate IT phase and the associated transition exhibit opposite uniaxial pressure dependencies in-plane and out-of-plane. In addition, the IT phase also shows a small but distinct non-volume-conserving temperature dependence of the Gr\"{u}neisen parameters (see Fig.~\ref{gruen}). This again indicates the presence of several degrees of freedom and may be interpreted as signature of a slowly evolving long- or short-range orbital order. Accordingly, the fluctuation regime below 10~K may indicate the presence of short-range orbital order competing with cAFM-type of magnetism.

Finally, we note finite values of $\delta$ up to about 65~K, i.e., the evolution of anisotropy in the length changes as also seen in Fig.~\ref{fig1}, indicating short-range fluctuations in this temperature regime. As visible in the inset of Fig.~\ref{fig1}, changes in $\delta$ above \tn\ are associated with anomalous in-plane expansion which yields a negative thermal expansion coefficient $\alpha_{\rm [110]}$ (cf. Fig.~\ref{alphai}). As both magnetic and quadrupolar degrees of freedom are strongly coupled to the lattice, this phenomenon may be in general related to fluctuations of either ordering phenomena. It is however not directly associated with fluctuations of any of the long-range ordering phenomena actually observed at low temperatures so that its nature remains unclear. Specifically, the uniaxial Gr\"{u}neisen parameter $\gamma_{\perp c}$ displays a sign change at \tn\ which implies opposite pressure dependencies $\partial \ln{\epsilon}/\partial p_c$ above and below \tn , thereby excluding the short-range fluctuations being of cAFM nature. An analogous argument holds for the LT-type  order. We also exclude that precursing anisotropy above \tn\ is associated with order of the IT-type as $\delta$ is suppressed and $L_{[110]}$ considerably shrinks in this phase. 

\section{Summary}

In summary, high-resolution capacitance dilatometry studies on the uniaxial length changes in a single crystal enable to elucidate the interplay of spin and quadrupolar order in NdB$_4$. The data imply significant magneto-elastic coupling and a structural phase transition at \titwo . While cAFM and LT favor structural anisotropy between in-plane and out-of-plane length changes, ${\delta = (dL_{[110]}-dL_{[001]})/(dL_{[110]}+dL_{[001]})}$ competes with order in the IT phase. Notably, finite anisotropy well above \tn\ indicates short-range correlations which are however of neither cAFM, IT, nor LT-type. Gr\"{u}neisen analysis is used to derive the uniaxial as well as the hydrostatic pressure dependencies. While the ratio $\alpha$/\cp\ features a single dominant energy scale in LT, the data implies precursory fluctuations of a competing phase in IT and cAFM. Our results suggest the presence of orbital degrees of freedom competing with cAFM and evolving towards the magnetically and orbitally ordered ground state.

\begin{acknowledgements}
Partial support by Deutsche Forschungsgemeinschaft (DFG) via Project KL 1824/13-1 and by BMBF via 
the SpinFun project (13XP5088) is gratefully acknowledged. SP acknowledges fellowship by the HGSFP. Work at the University of Warwick was funded by EPSRC (UK) through Grant EP/T005963/1.
\end{acknowledgements}

\bibliography{Bib_NdB4}

\end{document}